\documentclass[12pt]{article}
\date{}
\newenvironment{equation*}{%
  \equation
}{%
  \nonumber\endequation\nonumber
}
\def\ave#1{\langle #1 \rangle}
\def\pd{\partial}

\title{Relativity  and $c/\sqrt{3}$}
\author{S.I. Blinnikov, L.B. Okun, M.I. Vysotsky \\
   ITEP, 117218 Moscow, Russia }

\begin{document}
\maketitle

\begin{abstract}

We define the critical coordinate velocity ${\rm v}_c$. A particle
moving radially in Schwarz\-schild background with this velocity,
${\rm v}_c= c/\sqrt 3$, is neither accelerated, nor decelerated if
gravitational field is weak, $r_g \ll r$, where $r_g$ is the
gravitational radius, while $r$ is the current one.
We find that the numerical coincidence of ${\rm v}_c$  with velocity of sound
in ultrarelativistic plasma, $u_s$, is accidental, since two velocities
are different if the number of spatial dimensions is not equal to 3. 

\end{abstract}

{\bf To Holger Nielsen}

  We dedicate this note to our friend Holger Nielsen on the occasion 
of his 60th birthday. It  has been a great pleasure to discuss with
him exiting physical ideas at ITEP, CERN, the Niels Bohr Institute
and in other places around the world. We wish to Holger many new
discoveries and a long happy life.

\section{Motivation}

According to General Relativity (GR) clocks run slowly in the
presence of gravitational field, as a result, the coordinate
velocity of photons decreases. This is the reason for the delay of
radar echo from inner planets predicted and measured by I. Shapiro
\cite{1}. Propagation of ultrarelativistic particles is described
similarly to that of photons. That is why the retardation must
take place not only for photons but also for ultrarelativistic
particles. In this respect the latter drastically differ \cite{2}
from nonrelativistic bodies, velocity of which evidently increases
when they are falling radially onto a gravitating body (e.g., onto
the Sun). Obviously, there should be some intermediate velocity
${\rm v}_c$ which remains constant for a particle falling in
gravitational field of the Sun (or another star). The numerical
value ${\rm v}_c = c/\sqrt 3$ will be found in Sect. 2. When a
particle moves radially with this velocity in weak field it
``ignores'' gravity: it is neither accelerated, nor decelerated.
For nonradial trajectories gravity is never ignored: the
trajectories are bent for any velocity.

It is well known that $u_s = c/\sqrt 3$ is the speed of sound in
ultrarelativistic plasma and the question arises whether the
equality $u_s = {\rm v}_c$ has some physical reason, or it is a
numerical coincidence. To answer this question we find in Sect. 3
expressions for ${\rm v}_c$ and $u_s$ in spaces with number of
dimensions $n$ different from 3. Since for $n\neq 3$ we get ${\rm v}_c
\neq u_s$ we come to the conclusion that their coincidence at
$n=3$ does not have deep physical reason.

\section{Derivation of $\mbox{\boldmath${\rm v}_c = c/\sqrt 3$}$}

To simplify formulas, we put light velocity $c=1$, restoring it
when it is necessary. In what follows $G$ is gravitational constant;
gravitational radius $r_g$ of an object with mass $M$ equals
\begin{equation}
  r_g=2GM \; .
\label{rg}
\end{equation}

Let us start from definitions used in GR. The expression for
interval in the case of radial motion ($d\theta = d\varphi = 0$)
has the well known  Schwarzschild form:
\begin{equation}
ds^2 = g_{00}dt^2 -g_{rr}dr^2 \equiv d\tau^2 -dl^2 \;\; ,
\label{1}
\end{equation}
where $g_{00} =(g_{rr})^{-1} =1-\frac{r_g}{r}$.
The local velocity $v$ of a particle measured by a local observer
at rest is:
\begin{equation}
v = \frac{dl}{d\tau} =\left(\frac{g_{rr}}{g_{00}}\right)^{1/2}
\frac{dr}{dt}= \frac{1}{g_{00}} \frac{dr}{dt} \;\; , \label{2}
\end{equation}
while observer at infinity, where $g_{00}(\infty) =
g_{rr}(\infty) =1$, measures the so-called coordinate velocity at $r$:
\begin{equation}
{\rm v} = \frac{dr}{dt} =
v\left(\frac{g_{00}}{g_{rr}}\right)^{1/2} = g_{00}v \;\; .
\label{3}
\end{equation}
In order to determine the time of radial motion from $a$ to $b$,
the infinitely distant observer should calculate the integral
\begin{equation}
t =\int_a^b\frac{dr}{{\rm v}} \;\; , \label{4}
\end{equation}
that is why the coordinate velocity is relevant for radar echo.

For a particle moving in static gravitational field one can
introduce conserved energy (see ref. \cite{3}, eq. 88.9):
\begin{equation}
E =\frac{m\sqrt{g_{00}}}{\sqrt{1- v^2}} \;\; .
\label{5}
\end{equation}
The expression for $E$ through ${\rm v}$:
\begin{equation}
E
=\frac{m\sqrt{g_{00}}}{\sqrt{1-({\rm v}/g_{00})^2}}
\label{6}
\end{equation}
allows us to determine ${\rm v}(r)$ from the energy conservation:
\begin{equation}
E(r = \infty) =E(r) \;\; , \label{7}
\end{equation}
\begin{equation}
{\rm v}^2 = g_{00}^2 -g_{00}^3 +g_{00}^3 {\rm v}_\infty^2 =
g_{00}^2 [1-g_{00}(1-{\rm v}_\infty^2)] \;\; . \label{8}
\end{equation}
For the local velocity $v$ measured by a local observer
we obtain:
\begin{equation}
v^2 = 1-g_{00}(1-{\rm v}_\infty^2) \;\; , \label{9}
\end{equation}
so, while $v$ always increases for a falling massive particle, reaching $c$ at
$r = r_g$, the behaviour of ${\rm v}$ is more complicated. Substituting
$g_{00} =1-\frac{r_g}{r}$ into (\ref{8}), we get for weak gravitational field
($r\gg r_g$):
\begin{equation}
{\rm v}^2 = {\rm v}_\infty^2 +\frac{r_g}{r}(1-3{\rm v}_\infty^2)
\;\; . \label{10}
\end{equation}
For the motion of a nonrelativistic particle (${\rm v}_\infty \ll 1$) the well-known
expression is reproduced:
\begin{equation}
{\rm v}^2 = {\rm v}_\infty^2 +\frac{2MG}{r} \;\; .
\label{11}
\end{equation}
For ${\rm v}_\infty = {\rm v}_c = 1/\sqrt 3$ the coordinate velocity
of particle does not change, while it grows for ${\rm v}_\infty <
{\rm v}_c$ and diminishes  for ${\rm v}_\infty > {\rm v}_c$.
At $r = 3r_g$ according to eq.(\ref{10}) the coordinate velocity becomes equal to ${\rm v}_c$.
However, for $r=3r_g$ our weak field approximation fails.

Let us dispose of the assumption of the weak field.
Coming back to expression (\ref{8}) and substituting there $g_{00}
=1-\frac{r_g}{r}$, we observe that for ${\rm v}_\infty > {\rm v}_c$
the coordinate velocity always diminishes and becomes zero at $r=r_g$,
while in the case ${\rm v}_\infty < {\rm v}_c$ it grows up to the
value
\begin{equation}
{\rm v}_{\rm max}^2 =4/(27(1-{\rm v}_\infty^2)^2) \; ,
\label{vmax}
\end{equation}
which
is reached at
\begin{equation}
r_0 = \frac{3(1-{\rm v}_\infty^2)}{(1-3{\rm v}_\infty^2)}r_g \; ,
\label{r0}
\end{equation}
and after that diminishes to zero at $r=r_g$.
It is interesting to note that the velocity $v$ measured by local
observer equals ${\rm v}_c$ at the point where
${\rm v} = {\rm v}_{\rm max}$.

Thus, if the coordinate velocity is only mildly relativistic,
${\rm v}_\infty > c/\sqrt 3$, then ${\rm v}$ already decreases at
the free fall.

As an example of a non-radial motion let us consider the deflection
of light from a star by the Sun and compare it with the deflection of a
massive particle.
It is well known that the angle of deflection $\theta$ of photons
grazing the Sun is given by
\begin{equation}
 \theta_\gamma=\frac{2r_g}{R_\odot}
\label{thetagam}
\end{equation}
where $R_\odot$ is the radius of the Sun.
In the case of massive particles the deflection angle is larger:
\begin{equation}
  \theta = \theta_\gamma(1 + \beta^{-2}) \; ,
\label{theta}
\end{equation}
where  $\beta \equiv v_\infty/c < 1$.(See ref.\cite{MTW}, eq.
25.49, and ref.\cite{lightm}, problem 15.9, eq.13.)

\section{Speed of sound $u_s$ and critical speed
  v$_c$  in $n$ dimensions}

For ultrarelativistic plasma with equation of state $P=e/3$, where
$P$ is pressure and $e$ is energy density (including mass density),
we have for the speed of sound $u_s$:
\begin{equation}
 u_s^2 =c^2 \left.\frac{\partial P}{\partial e}\right|_{\rm ad} =
\frac{c^2}{3} \;\; , \label{12}
\end{equation}
We use eq. 134.14 of ref.  \cite{4}, and correct misprint there, or
eq. 126.9 from \cite{lanliffm}; ``ad'' means adiabatic, i.e. for constant specific
entropy.
In order to obtain the expression for $u_s$ in the case when $n\neq 3$,
where $n$ is the number of spatial dimensions, let
us start with equation of state.

One can use a virial theorem to connect pressure $P$ and thermal
energy $\mathcal{E}$ of an ideal gas using classical equations of
particle motion (cf. \cite{6}). We have for a particle with
momentum ${\rm\bf p}$ and a Hamiltonian $H$:
\begin{equation}
 \dot{\rm\bf p} = - \frac{\pd H}{\pd {\rm\bf q}} \; ,
\label{13}
\end{equation}
hence,
\begin{equation}
 {\rm\bf q} \dot{\rm\bf p} = - {\rm\bf q} \frac{\pd H}{\pd {\rm\bf q}} =
 {\rm\bf q} {\rm\bf F} \;\; ,
\label{14}
\end{equation}
where ${\rm\bf F}$ is the force acting on the particle.
Let us average over time $t$:
\begin{equation}
 \ave{\dots} \equiv {1\over t} \int_0^t \dots d\bar t \;\; .
\label{15}
\end{equation}
Integrating  by parts we get:
\begin{equation}
   \ave{ {\rm\bf q} \dot{\rm\bf p}} = -  \ave{\dot{\rm\bf q} {\rm\bf p}} =
   \ave{ {\rm\bf q} {\rm\bf F}} \;\; .
\label{16}
\end{equation}
For non-relativistic (NR) particles
\begin{equation}
  \dot {\rm\bf q} {\rm\bf p} = 2 E_{\rm kin} = {\rm\bf p}^2/m \; .
\label{17}
\end{equation}
For extremely relativistic (ER) particles
\begin{equation}
  \dot{\rm\bf q} {\rm\bf p} =  E_{\rm kin} = c |{\rm\bf p}| \; .
\label{18}
\end{equation}
Now for $N$ particles in a gas
\begin{equation}
   - \sum_{i=1}^N   \ave{   \dot{\rm\bf q}_i {\rm\bf p}_i } =
       \sum_{i=1}^N   \ave{ {\rm\bf q}_i {\rm\bf F}_i }.
\label{19}
\end{equation}
(By the way, $-\frac{1}{2}\sum\limits_i \ave{ {\rm\bf q}_i {\rm\bf
F}_i }$ is called the {\em virial}.) If the gas is ideal (i.e.
non-interacting particles), then the force ${\rm\bf F}$ is
non-zero only at the collision of a particle with the wall, and
the virial reduces to an integral involving pressure:
\begin{equation}
 - \sum_{i=1}^N   \ave{   \dot{\rm\bf q} {\rm\bf p}}
      = - \int P \mathbf{n} {\rm\bf q} \, dS =
   -P\int \mbox{div}\, {\rm\bf q} \, dV = -3PV \; ,
\label{20}
\end{equation}
where $\mathbf{n}$ is a unit vector normal to the wall area element $dS$ and
the Gauss theorem is used for transforming the surface integral to the volume one.
So, since the thermal energy $\mathcal{E}$  (not including mass) is just
the total kinetic energy of molecules,
\begin{equation}
  \mbox{NR} : \quad 2\mathcal{E} = 3PV, \quad P = 2\mathcal{E}/(3V)  \; ,
\label{21}
\end{equation}
\begin{equation}
  \mbox{ER} : \quad \mathcal{E} = 3PV,  \quad P =  \mathcal{E}/(3V) \equiv e/3 \; .
\label{22}
\end{equation}

The last equality holds since in extremely relativistic case
$E_{\rm kin} \gg m$. We see that $3$ here is due to $\mbox{div}\,
{\rm\bf q}=3$, i.e. the dimension of our space.

In a space of $n$ dimensions, following the same lines, we get
$\mbox{div}\, {\rm\bf q}=n$, so $P=e/n$ and for ER gas we obtain:
\begin{equation}
u_s = c/\sqrt n \;\; .
\label{23}
\end{equation}
 Here we should use
$n$-volume $V_n$ instead of $V\equiv V_3$ and postulate the first
law of thermodynamics for adiabatic processes to be $d\mathcal{E}+PdV_n=0$,
so pressure would be the force per unit $V_{n-1}$ -- the boundary
of $V_n$.

The same equation of state follows from consideration of the
stress tensor $T_{ik}$ of ultrarelativistic plasma, which is diagonal and
traceless in a rest frame of plasma: $T_{00} = e$, $T_{ii} \equiv P = e/n$.

In order to find ${\rm v}_c$ in the case $n\neq 3$ we need an
$n+1$-dimensional spherically-symmetric static generalization of the
$3+1$-dimensional Schwarz\-schild metric which was found by
Tangherlini \cite{tang}. (See refs.\cite{MP} for the details
of aspherical and time-dependent black holes metrics in higher
dimensional spacetimes). The line element of the $n+1$-dimensional
Schwarz\-schild metric is
\begin{equation}
 ds^2 = \left(1 - {r_{gn}^{n-2} \over r^{n-2}}\right) dt^2 -
 \left( 1 - {r_{gn}^{n-2} \over r^{n-2}}\right)^{\!-1} dr^2 - r^2 d\Omega_{n-1}^2,
\label{Nsch}
\end{equation}
where $d\Omega_{(n-1)}$ is the line element on the unit
$(n-1)$-sphere
and the gravitational radius $r_{gn}$ is related to the black hole mass $M$:
\begin{equation}
r_{gn}^{n-2} = {16 \pi G_n M \over (n - 1) \ A_{n-1}}.
\label{rgn}
\end{equation}
Here $A_{n-1}$ denotes the area of a unit $n-1$ sphere, which
is ${{2 \pi^{n \over 2}} / \Gamma({n \over 2})}$ (for $n=3$:
$\Gamma(3/2)=\sqrt{\pi}/2$, $A_2=4\pi$).
We consider the spaces with $n \geq 3$.
The factor in the definition of $r_{gn}$
is taken from refs.\cite{MP}. 
The form of the metric (\ref{Nsch}) is very easy to guess.
In weak field  limit, when $g_{00} \to 1+2\varphi $
we have
\begin{equation}
   \varphi = - \frac{r_{gn}^{n-2}}{2r^{n-2}}
  \quad \mbox{for} \quad r \to \infty \; .
\label{pot}
\end{equation}
This leads in a natural way to the gravitational acceleration $\mathbf{g}$
with the radial component
\begin{equation}
   g_n = -\frac{\pd \varphi}{\pd r}= - \frac{(n-2) r_{gn}^{n-2}}{2r^{n-1}}
  \quad \mbox{for} \quad  r \to \infty \; ,
\label{gacc}
\end{equation}
which implies the constant flux of the acceleration $\mathbf{g}$
equal to
\begin{equation}
 A_{n-1}r^{n-1}g_n = 8\pi \frac{n-2}{n-1}  G_n M
\label{fluxg}
\end{equation}
through a sphere of area $A_{n-1}r^{n-1}$ at large $r$.
It is not hard to verify that the Ricci tensor $R_{ik}$ is zero for the
metric (\ref{Nsch}), that is the metric (\ref{Nsch}) satisfies Einstein
equations in vacuum and describes a spherically symmetric spacetime outside
a spherical gravitating body.

One should remember that the dimension of $G_n$ depends on $n$.
It is clear in the weak field limit from eqs. (\ref{pot}) and (\ref{rgn}),
since
the dimension       
of $[\varphi]$ is the square of velocity, i.e. zero for $c=1$, and hence
\begin{equation}
   [G_n] = {L^{n-2} \over M} \; ,
\label{27}
\end{equation}
or $[G_n] = L^{n-1}$, if $[M]=L^{-1}$.

For the coordinate velocity of a radially falling  particle we get from eq.
(\ref{8}) for weak gravitational field:
\begin{equation}
{\rm v}^2 = {\rm v}_\infty^2 +\left(\frac{r_{gn}}{r}\right)^{n-2}
                     (1-3{\rm v}_\infty^2) \label{28}
\end{equation}
instead of eq. (\ref{10}). We see that in the case of
$n$-dimensional  space the expression for ${\rm v}_c$ remains the
same, ${\rm v}_c =c/\sqrt 3$. Number ``3'' here is not due to the
dimension of space, it is  simply due to cubic polynomial in
(\ref{8}).

\section{Conclusions and Acknowledgements}

The speed of sound in relativistic, radiation dominated plasma
depends on the dimension  of space, while the critical velocity
${\rm v}_c=c/\sqrt{3}$ in the Schwarzschild metric is the same for
any dimension.

\bigskip
We are grateful to Ilya Tipunin for encouraging our search of
literature on black hole metrics in higher dimensions and to Yakov
Granowsky and Valentine Telegdi for their interest to the
discussed problem and stimulating questions. SB is partially
supported by RFBR grant 99-02-16205, and by a NSF grant at UCSC,
he is especially grateful to Stan Woosley for his hospitality. LO
is supported by A.v.Humboldt award and together with MV partly
supported by RFBR grant No. 00-15-96562.

PS. After the first version of this paper appeared on the web we
recieved an e-mail from Stanley Deser and Bayram Tekin in which
it was pointed out that critical velocity in weak field has been
considered by M.Carmeli: Lettere al Nuovo Cimento 3 (1972)379
and in his book ``Classical Fields: General Relativity and
Gauge Theory'', John Wiley and Sons, Inc 1982 New York . We thank them for 
this bitter remark.

\end{document}